\documentclass[twocolumn,superscriptaddress,aps,prb]{revtex4-1}
\usepackage{graphicx}
\usepackage{dcolumn}
\usepackage{bm}
\usepackage{color,graphicx}
\usepackage{amsmath}
\usepackage{eqnarray}
\usepackage{amssymb}
\usepackage{amsthm}
\usepackage{amsfonts}
\usepackage{braket}
\usepackage[arrowdel]{physics}
\pdfoutput=1 
\begin{document}

\title{Beyond equivalent circuit representations in nonlinear systems with inherent memory}


\author{Victor Lopez-Richard}
\affiliation{Departamento de Física, Universidade Federal de São Carlos, 13565-905 São Carlos, SP, Brazil}
\email{vlopez@df.ufscar.br}

\author{Soumen Pradhan}
\affiliation{Julius-Maximilians-Universität Würzburg, Physikalisches Institut and Würzburg-Dresden Cluster of Excellence ct.qmat, Lehrstuhl für Technische Physik, Am Hubland, 97074 Würzburg, Deutschland}
\email{soumen.pradhan@uni-wuerzburg.de}

\author{Rafael Schio Wengenroth Silva}
\affiliation{Departamento de Física, Universidade Federal de São Carlos, 13565-905 São Carlos, SP, Brazil}

\author{Ovidiu Lipan}
\affiliation{Department of Physics, University of Richmond, 28 Westhampton Way, Richmond, Virginia 23173, USA}

\author{Leonardo K. Castelano}
\affiliation{Departamento de Física, Universidade Federal de São Carlos, 13565-905 São Carlos, SP, Brazil}

\author{Sven Höfling}
\affiliation{Julius-Maximilians-Universität Würzburg, Physikalisches Institut and Würzburg-Dresden Cluster of Excellence ct.qmat, Lehrstuhl für Technische Physik, Am Hubland, 97074 Würzburg, Deutschland}

\author{Fabian Hartmann}
\affiliation{Julius-Maximilians-Universität Würzburg, Physikalisches Institut and Würzburg-Dresden Cluster of Excellence ct.qmat, Lehrstuhl für Technische Physik, Am Hubland, 97074 Würzburg, Deutschland}

\date{\today}

\begin{abstract}
Basic multimode impedance analysis, based on the availability of nonequilibrium charge carriers and their delayed return to equilibrium, is employed to assess the state of equivalent circuit representations. This analysis highlights the necessity of surpassing these representations in nonlinear systems with inherent memory, along with their associated advantages and limitations. On the basic grounds of generation and recombination (or trapping) of nonequilibrium carriers and their relaxation times, we show how seeming complexity of frequency-dependent impedance that matches a vast universe of experimental evidences can be reduced to simple combinations of basic microscopic ingredients. Counterintuitive features, such as a negative capacitances or unexpected inductances, arise when the results are projected onto linear equivalent circuit representations. 
This indicates the presence of certain limitations and potential ambiguities in the symbolic representation of 'equivalent' circuits. Our approach further provides a microscopic perspective that exposes the linkage of an apparent flux with an apparent inductance dismissing any magnetic essence.
\end{abstract}

                              
\maketitle

\section[here]{Introduction}

Be it intentional or not, the nonlinear response of conductive systems or the nonlinear operation of electronic devices are widely present. It appears in redox reactions at the surface of metamaterials~\cite{Priyadarshani2021}, leaking currents in solar cells~\cite{Guerrero2016}, memristive effects in the conduction of semiconductor oxides~\cite{Paiva2022}, electrochemical processes within energy storage components~\cite{Luo2022}, memory functionalities of nanoscopic transistors~\cite{Miller2021,Bisquert2023}, among many other systems. In frequency domain, these nonlinear effects are characterized by combining cyclic voltammetry and impedance spectroscopy (IS)~\cite{Li2013,Shangshang2021,Ebadi2019,Priyadarshani2021,Gonzales2022} with the purpose to understand their driving mechanisms while quantifying the ruling parameters. The latter consists in the Fourier expansion of the system transport stable response to sinusoidal drives. When the transport response deviates from linearity, a set of challenges emerges~\cite{Kowal2009} such as the impedance dependence on the pulse amplitude~\cite{Fasmin2017} and higher mode generation~\cite{Kiel2008,Pershin2021}. Both are linked to the microscopic carrier dynamics, which is an often overlooked connection. Another challenge is the effort to translate the nonlinear response in terms of combinations of ``equivalent'' linear circuit elements and subsequently trying to ascribe physical meaning to them. Inconsistencies in interpreting IS results~\cite{Mei2018} may require reevaluating certain behaviors, including seemingly inductive effects~\cite{Bisquert2024,Klotz2019,Bou2020,Abdulrahim2021,Priyadarshani2021,Gonzales2022,Thapa2022}, often described as negative capacitance~\cite{Jonscher1986,Ershov1998,MoraSero2006,Ebadi2019,Joshi2020}, and some apparent capacitive responses possibly unrelated to electric displacement~\cite{Irvine1990,Liu2005,Das2008,Li2013,Nowroozi2018,Vadhva2021}.

Equivalent circuits offer a simplified and intuitive way to model complex systems, making it easier to understand and predict device behavior. They can be integrated into circuit simulation software to model the overall behavior of electronic systems, facilitating system-level design and optimization. Under certain operating conditions, these models provide reasonable approximations, making them valuable for initial design, testing, and troubleshooting. Equivalent circuits also simplify the interpretation of complex impedance data by representing it through familiar components such as resistors, capacitors, and inductors~\cite{Bisquert2024a,Bisquert2024b,Gonzales2024}. They offer insights into frequency-dependent behaviors like reactance and resonance, enabling efficient analysis. Once established, equivalent circuits can predict device impedance across different conditions, reducing the need for time-consuming calculations. Additionally, they serve as diagnostic tools by identifying deviations in performance and isolating problematic components. By providing a standardized framework, equivalent circuits allow for consistent comparison of impedance characteristics across different devices and materials.

In this work, we show how the natural dynamics of nanoscopic nonlinear systems with inherent memory highlight some limitations of equivalent circuit representations and reveal how mapping these dynamic elements onto steady-state classical passive circuit components can lead to unexpected responses. To meet these challenges, we take advantage of the multimode characterization of nonlinearities~\cite{Fasmin2017} in combination with a microscopic perspective of processes driven out of equilibrium. We show how the interplay between competing intrinsic timescales and symmetry of generation functions of coexisting independent transport channels impose the need to perform a characterization beyond the fundamental mode.

\begin{figure}[h]
	\includegraphics[width=8.5cm]{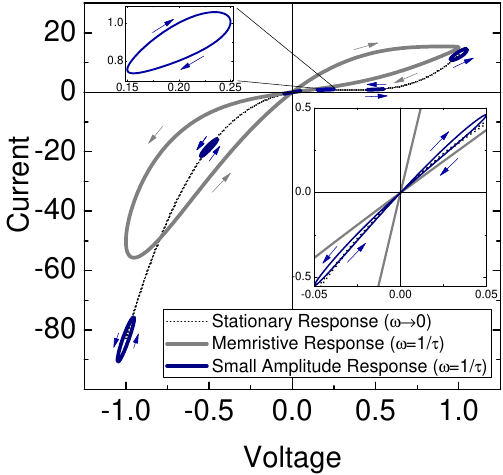}
	\caption{\label{Figure1} Theoretical simulation of nonlinear dynamic response. Current-voltage response of a device with memory: memory response under AC voltage drive (grey line), stationary response (dotted line), and under a stationary DC bias and small amplitude AC drive (blue curves). The insets show a zoom close to zero and 0.2 voltages. The loop directions are indicated by arrows.}
\end{figure}

\begin{figure}[h]
	\includegraphics[width=8.5cm]{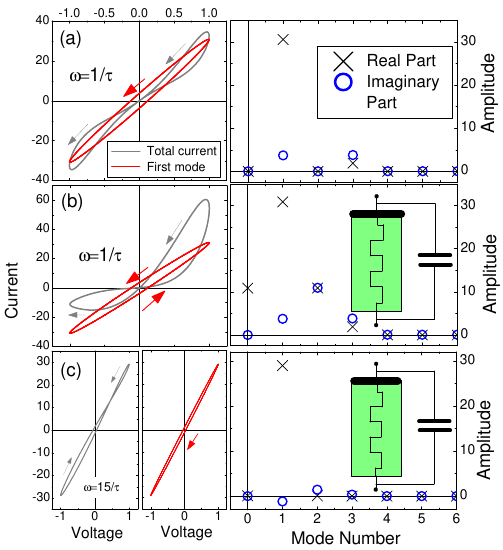}
	\caption{\label{Figure2new} \textbf{Theoretical simulation of the effect of symmetry and frequency in nonlinear dynamic response.} (a) Left:  Symmetric current-voltage memory response (grey curve) and the its corresponding first mode contribution (red ellipse) and right: multimode Fourier expansion of the current amplitudes. (b)  Left: Asymmetric current-voltage memory response (grey curve) in the presence of a capacitive coupling between contacts and the its corresponding first mode contribution (red ellipse) and right: multimode Fourier expansion of the current amplitudes. (c) Effect of increasing the frequency 15 times in the response depicted in panel (b). The insets correspond to a memristive device with a capacitive coupling used for the two last panels. The loops directions are indicated by arrows.}
\end{figure}

\section[here]{Discussion}

The generation or trapping of nonequilibrium carriers affects the conductance of nanoscopic devices and can originate from various (independent) nonequilibrium generation/trapping mechanisms e.g., electrons trapped in deep or shallow defects~\cite{Paiva2022}, the diffusion of ionic species, or the contributions of different chemical compounds adsorbed on a surface~\cite{Messerschmitt2015}. Within the relaxation time approximation, the carrier concentration fluctuation, $\delta n$, towards equilibrium can be written as
\begin{equation}
    \label{taxa}
    \frac{d \delta n}{dt}=-\frac{\delta n}{\tau} + g\left( V \right),
\end{equation}
or as its integral form
\begin{equation}
    \label{taxa2}
    \delta n(t)=\int_0^t g(t-t') e^{-\frac{t'}{\tau}} dt' + \delta n(0) e^{-\frac{t}{\tau}},
\end{equation}
with $\tau$ being the relaxation time and $\delta n(0)$, the initial condition. The generation and trapping rates are described by the function $g\left( V \right)$. It contains microscopic information on the nature of these processes such as thermal activation, quantum tunneling, etc., along with the dependence on external parameters, e.g. electric field and temperature. It is worth noting that, independently on the initial conditions in Eq.~\ref{taxa2}, a stable solution can be reached for after a transient forming process, by applying $V=V_{0} \cos{\omega t}$ and waiting long ``enough''~\cite{Paiva2022,Lopez-Richard2024b} for the transient terms on the right-hand side to vanish. 
Then, nonlinear conductance emerging from the state dynamics in Eqs.~\ref{taxa} or~\ref{taxa2} 
can be emulated using the Drude like conductance~\cite{Silva2022}, where the nonequilibrium carriers, $\delta n$, contribute as 
\begin{equation}
    \label{curr}
    I= \left( G_{0} +  \gamma \delta n \right) V,
\end{equation}
according to their mobility, $\mu$, with $\gamma=e \mu/ d^2$ and $d$ being the distance between contacts~\cite{Paiva2022}, while $G_{0}$ is the unperturbed conductance. 

For small relative signal amplitudes, as those used in linear IS, the generation/trapping rate can be decomposed up to second order on the applied voltage, 
\begin{equation}
    \label{gen}
g\left( V \right) \simeq \sigma_{o} V+\sigma_{e}V^2. 
\end{equation}
All the microscopic information resides within $\sigma_{o}$ and $\sigma_{e}$ while their sign, relative strength, along with the voltage polarity (relevant for the $\sigma_{o}$ term) define whether $g<0$, for a trapping or, $g>0$ for the generation character of the sites of nonequilibrium carriers. Under a  time-periodic voltage, $V=V_{0} \cos{\omega t}$, the presence of generation or trapping phenomena induces a hysteresis behavior in the current-voltage plane~\cite{LopezRichard2022}, as depicted in Figure~\ref{Figure1}. It displays stable cycles characterized by asymmetric loops with $\sigma_{o}<0$ and $\sigma_{e} \ne 0$, exhibiting a crossing at $V=0$, which are identified as Type I memristive responses. In turn, the left panel Figure~\ref{Figure2new} (a) showcases a symmetric stable loop obtained when $\sigma_{o}=0$, depicting a pinched hysteresis without a crossing at $V=0$, labeled as Type II response~\cite{Pershin2011,Silva2022}.
Additional loop geometries can be found in Ref.~\citenum{LopezRichard2022}. 
The picture can be further expanded to incorporate a geometric capacitive coupling, $C>0$, whether intentional or parasitic, by introducing the term, $-C \omega V_{0} \sin{\omega t}$, to Eq.~\ref{curr}. In the asymmetric responses depicted in Figures~\ref{Figure2new} (b) and (c) $C>0$ has been considered. It is discernible as openings in the current-voltage loops at $V=0$, that become pronounce with increasing frequency, as shown in the left panel of Figure~\ref{Figure2new} (c). In this instance, $\sigma_o >0$ has been used to highlight the contrast with Figure~\ref{Figure1}, which arises from the sign of the generation function terms affecting the direction of the loop.

As highlighted in Figure~\ref{Figure1}, the clockwise direction of the memory response loop at positive polarity is not mirrored by the small loops, except at sufficiently low DC voltages. Preventing these discrepancies is one reason we recommend employing large-amplitude spectral analysis, as it provides a more accurate depiction of the memory response characteristics and the correlation between current-voltage loops and impedance mapping.

Note that the opposite sign of the term $\sigma_o$ in the microscopic generation function is responsible for reversing the direction of the hysteresis loops in Figure~\ref{Figure1}, where $\sigma_o < 0$, compared to Figure~\ref{Figure2new} (b) and (c), where $\sigma_o > 0$. This term makes the generation or trapping behavior of the function $g(V)$ polarity-dependent (for low enough bias). When $\sigma_o < 0$ dominates (e.g., at low amplitudes), the memory response manifests as a clockwise loop for positive polarity and a counterclockwise loop for negative voltages. Conversely, this behavior is inverted when $\sigma_o > 0$. However, as demonstrated below, this distinct behavior could not be captured by single-mode impedance analysis unless adequate DC bias is applied~\cite{Lopez-Richard2024} to make $\sigma_o$ dominate over the first order contribution of $\sigma_e$ terms.

For spectroscopic analysis to effectively utilize Fourier expansion, reasonably stable periodic responses are required. Thus, employing a small amplitude probe around a stationary DC bias, $V_s$, with $V=V_s+V_{0} \cos{\omega t}$, proves inadequate for assessing the volatile conductive states described in the nonlinear responses above. As depicted in Figure~\ref{Figure1}, the mismatch between the hysteresis loop and the stable small amplitude probe around a stationary bias is evident. Only by employing probing signals $V=V_s \cos{\omega_s t}+V_{0} \cos{\omega_0 t}$, with $\omega_0>>\omega_s \sim 1/\tau$, the probing states may briefly coincide with the memristive hysteresis loop~\cite{Kubicek2015,Gonzales2022}. Nevertheless, such high frequencies preclude the spectral analysis from capturing the underlying non-equilibrium mechanism responsible for the hysteresis phenomenon in the first place. Hence, we advocate for the utilization of either small or large amplitude spectroscopy always centered at $V=0$ as the optimal procedural approach in the IS in the presence of nonlinear dynamic responses.

\begin{figure}[t]
	\includegraphics[width=8.5cm]{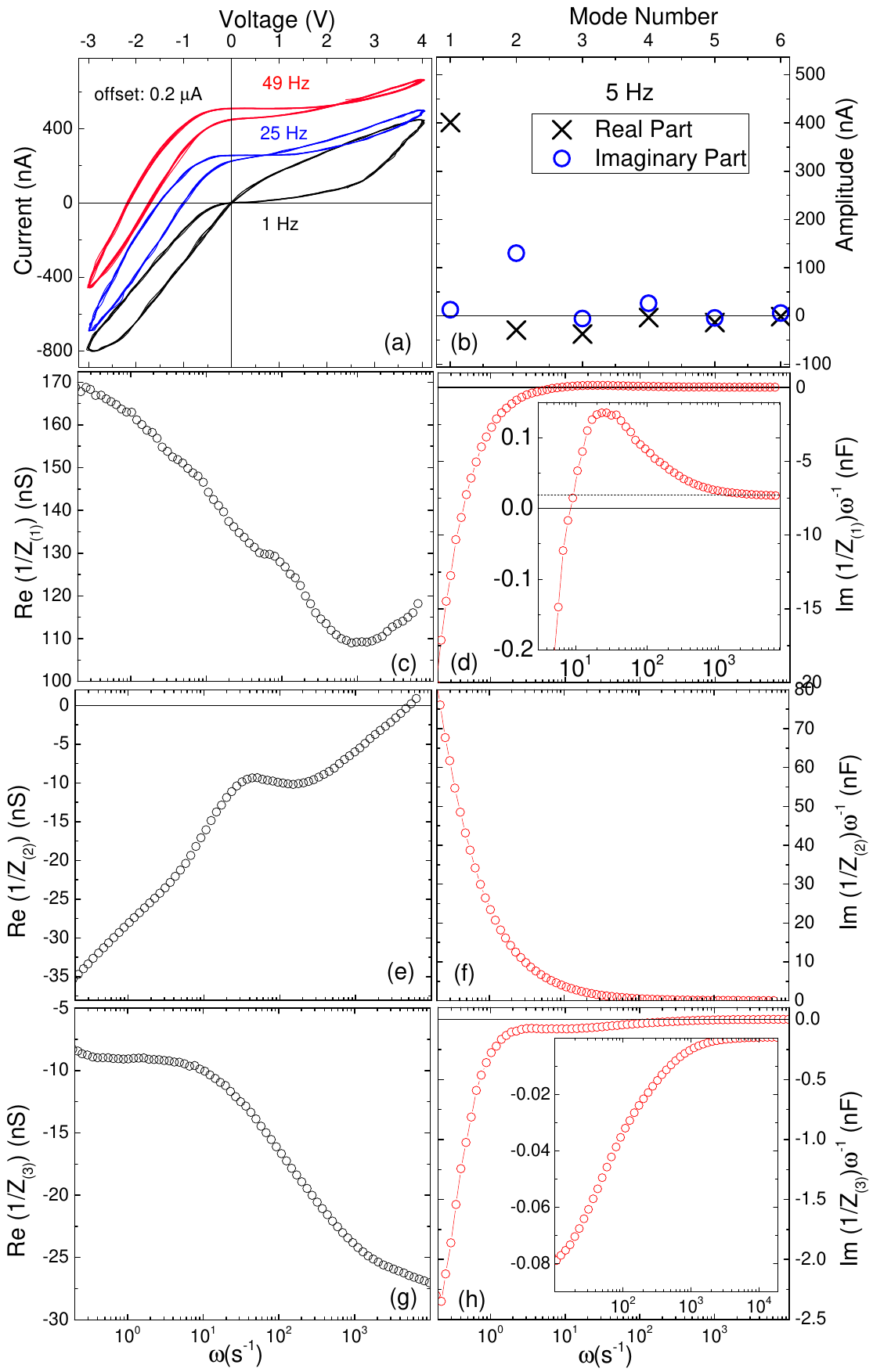}
	\caption{\label{Figure3} \textbf{Experimental multimode analysis of a memristive response.} (a) Current-voltage memristive loops at various frequencies of the AC drive. (b) Multimode Fourier expansion of the current amplitude at 5 Hz. Frequency dependence of $\text{Re}(1/Z_{(1)})$ and $\text{Im}(1/Z_{(1)})\omega^{-1}$, in panels (c) and (d). The inset shows a zoom into the large frequency range. The corresponding frequency tuning of $\text{Re}(1/Z_{(2)})$ and $\text{Im}(1/Z_{(2)})\omega^{-1}$, is shown in panels (e) and (f). While (g) and (h) display the values of $\text{Re}(1/Z_{(3)})$ and $\text{Im}(1/Z_{(3)})\omega^{-1}$, respectively.}
\end{figure}
 
Moreover, the corresponding Fourier coefficients for the current-voltage response, as detailed in the Methods Section, are presented on the right side of Figures~\ref{Figure2new} (a-c). The presence of higher Fourier modes in the response, compared to those in the driving voltage input, is inevitable in nonlinear dynamical systems, even for small amplitudes, as noted in previous studies~\cite{Kiel2008,Fasmin2017}. It is noteworthy that the pattern of the Fourier coefficients corresponds to the symmetry of the generation function: in Figure~\ref{Figure2new} (a), only the first and third modes are present, while the second mode becomes prominent for the asymmetric response in panel (b). These coefficients are influenced by the applied voltage amplitude and frequency, encapsulating the spectroscopic information. For instance, simply by increasing the frequency, the sign of the imaginary part of the first mode has changed from panel (b) to (c), resulting in the reversal of the ellipse direction.  

It is important to note that traditional IS analyses typically focus on characterizing the first-order mode~\cite{Kubicek2015,Bisquert2023a,Bisquert2023b}. Yet, the first-order contribution to each transport response inevitably corresponds to ellipses in the current-voltage plane (in presence or absence of $C$), which have been added as red curves into the left panels of Figures~\ref{Figure2new} (a-c). Note also that regardless of how small the amplitude is in the inset of Figure \ref{Figure1}, the current-voltage loop around $V=0$ will always exhibit a pinched hysteresis, which is distinct from the unavoidable ellipse of the fundamental mode. While these ellipses contain valuable information about the device response, the significant mismatch emphasizes the limitation of relying solely on first-order characterization. Therefore, to effectively characterize nonlinearities and memory emergence effects while retaining microscopic information, it is essential to explore beyond the first-order mode. An experimental confirmation of these trends has been included into Figure~\ref{Figure3}. The panel (a) displays a memristive response of a floating gate transistor~\cite{Miller2021} tuned by increasing frequency until a capacitive gap at $V=0$ becomes evident. 
The device in Fig.~\ref{Figure3} operates in 2-terminal memristive configuration, where the voltage was applied to one terminal and both the real and imaginary parts of the current were measured at other terminal with the voltage applied to the drain and the source grounded. Although there are lateral gate electrodes, these were kept floating during operation. The capacitance measured is the depletion capacitance of the device, which can be modulated by the applied drain voltage and its polarity.
The corresponding multimode expansion of the current response has been displayed in panel (b) for 5 Hz AC input. Then, subsequently, the spectroscopic frequency tuning of the complex admittance has been depicted in panels (c,d) for the first mode, (e,f) - second mode, and (g,f) - third mode, following the recipe described in the method section and Ref.~\citenum{Lopez-Richard2024b}. A seemingly complex picture emerges with either monotonic or nonmonotonic trends of the functions plotted. Note, for instance, that the $\text{Im} (1/Z_{(1)}) \omega^{-1}$ changes sign and attains a maximum before converging towards a finite positive value. Also, it is noteworthy that the functions of the second and third order modes exhibit step-like behavior in the logarithmic scale, albeit with contrasting monotonicity and sign. Thus, it is crucial to clarify the nature of information contained within each mode and to develop appropriate interpretations of spectroscopic analyses. Can we translate the complex qualitative and quantitative images of IS into fundamental microscopic components? Can we anticipate the kind of spectroscopy maps expected for nonlinear dynamic responses? Are auxiliary equivalent circuit representations helpful or not in comprehending and replicating these responses? These are the questions addressed in this paper.
 
We may start by assuming the relaxation time independent of the voltage amplitude. Thus,  Equations~\ref{taxa} and~\ref{curr} can be analytically solved resulting, up to second order in voltage, in
\begin{equation}
    \centering
    \label{curr2}
    \begin{aligned}
&I(t)=V_{0} \frac{A}{1+(\omega \tau)^2} - C\omega V_0\sin{\omega t}\\
& +V_{0}\left( G_0+\frac{S}{2} \right)\cos{\omega t} \\
&+\frac{V_0 S}{4}\left[\frac{\cos{\omega t}}{1+(2\omega \tau)^2}+\frac{2\omega \tau\sin{\omega t}}{1+(2\omega \tau)^2}   \right]  \\
& +V_0 A\left[\frac{\cos{2\omega t}}{1+(\omega \tau)^2}+\frac{\omega \tau\sin{2\omega t}}{1+(\omega \tau)^2}   \right] \\
&+\frac{V_0 S}{4}\left[\frac{\cos{3\omega t}}{1+(2\omega \tau)^2}+\frac{2\omega \tau\sin{3\omega t}}{1+(2\omega \tau)^2}   \right].
\end{aligned}
\end{equation}
Here, $S=\gamma\sigma_{e}\tau V_0^2$ and $A=\gamma\sigma_{o}\tau V_0/2$ are frequency independent parameters that depend on the applied voltage amplitude. The multimode expansion for small amplitudes, in Eq.~\ref{curr2}, stops at third order and has already been exemplified in Figures~\ref{Figure2new} (a-c). 

The contrasting asymmetric hysteresis shapes (with a crossing at zero bias) observed when the sign of the generation function parameter $\sigma_o$ changes, as shown in Figures~\ref{Figure1} and~\ref{Figure2new} (b), can only be captured through second-order impedance analysis, or for adequate finite DC bias.~\cite{Lopez-Richard2024} The $\sigma_o$ term specifically governs the change in carrier trapping or generation behavior based on voltage polarity. In contrast, first-order impedance analysis accounts solely for the contribution of the $\sigma_e$ term at zero DC bias, which are polarity-independent, according to the microscopic model up to the second-order approximation in Eq.~\ref{gen}. Therefore, accurately correlating apparent inductive or capacitive hysteresis loops~\cite{Bisquert2024} with corresponding impedance maps needs a thorough approach.

Besides the trivial geometric capacitive term, when $C\neq 0$ in Eq.~\ref{curr2}, the presence of dephasing proportional to $ \sin \omega t$, along all the modes can be interpreted as signatures of apparent reactive contributions. This points to the natural emergence of a dependence of the conductance on the \textit{apparent} flux, $\phi=\int_0^t V(t) dt=V_0/\omega \sin \omega t$, that has been a disputed missing connection for unambiguous definitions of memristors~\cite{Chua1971}. Then, as reported in  Ref.~\citenum{Strukov2008}, no real inductive effects are needed to explain such a link that emerges in our case from the simple and ubiquitous nonequilibrium dynamics.
\begin{figure}[t]
	\includegraphics[scale=1.0]{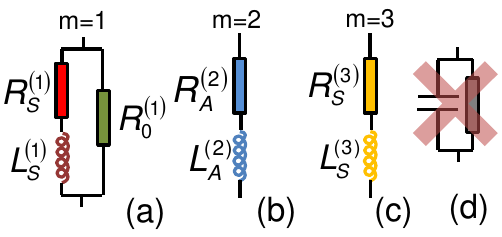}
	\caption{\label{Figure3new} \textbf{\textit{Apparent} circuits per mode.} (a), (b) and (c) \textit{apparent} circuits for the modes $m=1$, $m=2$, and $m=3$, respectively. (d) Not recommended $RC$ element. }
\end{figure}
The first line of Eq.~\ref{curr2} consists of a direct $m=0$ order contribution to the current, with an effective conductance, $1/R_{eff}^{(0)}=A/[1+(\omega \tau)^2]$, that arises from odd components to the generation function, available just in the absence of inversion symmetry, along with the geometric capacitive term proportional to $C$, that contributes to the fundamental mode, $m=1$. The second and third line also contribute to the fundamental mode resulting in an admittance for $m=1$, as defined in the Methods Section, given by
\begin{equation}
    \centering
    \label{adm1}
    \begin{aligned}
&\frac{1}{Z_{(1)}}=\left( G_0+\frac{S}{2} \right) \\
&+\frac{S}{4}\left[\frac{1}{1+(2\omega \tau)^2}-i \frac{2\omega \tau}{1+(2\omega \tau)^2}   \right] \\
&+iC\omega.
\end{aligned}
\end{equation}
The corresponding admittance of $m=2$ and $m=3$ is, respectively
\begin{equation}
    \label{adm2}
\frac{1}{Z_{(2)}}=A\left[\frac{1}{1+(\omega \tau)^2}-i \frac{\omega \tau}{1+(\omega \tau)^2}   \right],
\end{equation}
and
\begin{equation}
    \label{adm3}
\frac{1}{Z_{(3)}}=\frac{S}{4}\left[\frac{1}{1+(2\omega \tau)^2}-i \frac{2\omega \tau}{1+(2\omega \tau)^2}   \right].
\end{equation}
One may note the resistive contribution of the first line of Eq.~\ref{adm1} with an effective resistance $R_0^{(1)}=(G_0+S/2)^{-1}$ and the trivial geometric capacitive addition in the third if $C\neq 0$. In turn, given that $G_0$, $\tau$, $S$, and $A$ are frequency independent parameters, there is a common pattern in the functional dependence on $\omega$ of the second line of Eq.~\ref{adm1} and also in Eqs.~\ref{adm2} and~\ref{adm3}. We can thus draw an analogy between these cases and the admittance of a series $RL$ element (minimizing the number of circuit components) 
\begin{equation}
    \label{RL}
\frac{1}{Z_{RL}}=\frac{1}{R}\left[\frac{1}{1+(\omega \tau)^2}-i \frac{\omega \tau}{1+(\omega \tau)^2}   \right],
\end{equation}
with $\tau=L/R$. Thus, the apparent circuit representations of Eqs.~\ref{adm1},~\ref{adm2}, and~\ref{adm3} are provided in Figures~\ref{Figure3new} (a-c).
Due to the sign ambiguity of the generation function in Eq.~(\ref{gen}), the parameters $S$ and $A$ can assume either positive or negative values, resulting in different phase shifts between the real and imaginary components of the current. This behavior can be interpreted as inductive-like when the current lags the voltage, or capacitive-like when it leads.

There are various approaches to characterizing the processes of carrier release or trapping in response to changes in chemical potential, with the introduction of chemical capacitances being a well-established method.~\cite{Jamnik2001,Bisquert2022} In our work, we chose to correlate the microscopic description of the system’s ability to accept or release additional charge carriers with apparent inductive mechanisms (negative or positive) in our circuit models, ensuring that the elements remain independent of frequency. That is the reason we recommend avoiding the use of $RC$ elements, as represented in Figure~\ref{Figure3new} (d) with  $1/Z_{RC}=1/R+i \omega C$ to emulate the reactive responses in Eqs.~\ref{adm1}-\ref{adm3} as they inherently introduce a frequency dependence of the effective resistance and capacitance~\cite{Jonscher1986,Ershov1998,MoraSero2006,Ebadi2019,Joshi2020,Irvine1990,Liu2005,Das2008,Li2013,Nowroozi2018,Vadhva2021}. Treating the apparent circuit elements as frequency independent, whenever feasible, may be considered more advantageous for modeling.
However, an exception applies to diffusive components, which can deviate from the Warburg limit at low frequencies, as demonstrated in Ref~\citenum{Lopez-Richard2024}. These diffusive components behave in a capacitive-like manner, affecting current-voltage dephasing through an apparent capacitance that remains intricately tied to complex frequency dependencies, both in first-order and higher modes.

The use of \textit{apparent} circuits, as compact, symbolic, and intuitive representations of the system response can be useful and we have provided them for $m=1,2,3$ (setting $C=0$) in panels (a), (b), and (c) of Figure~\ref{Figure3new}, according to the functional dependence on frequency of the admittance obtained by the microscopic approach. Here, $R_S^{(1)}=4/S$, $L_S^{(1)}=8\tau/S$, $R_A^{(2)}=1/A$, $L_A^{(2)}=\tau/(2A)$, $R_S^{(3)}=R_S^{(1)}$, and $L_S^{(3)}=8\tau/(3S)$. 
Although being independent of frequency, these effective parameters depend on the input amplitude and thus cannot be interpreted as intrinsic properties decoupled from the external drive. Note also that it is expected for them to naturally change sign according to the sort of the generation function components ($\sigma_{o}$ or $\sigma_{e}$), as stated previously. 

\begin{figure}[t]
	\includegraphics[scale=1.0]{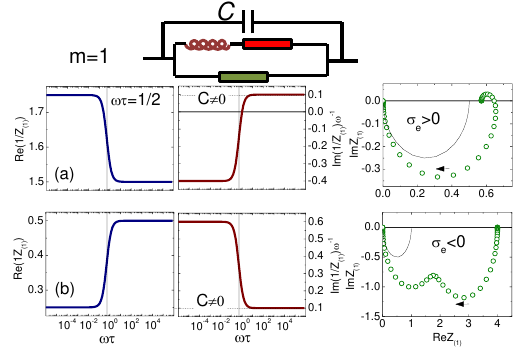}
	\caption{\label{Figure4} \textbf{Theoretical impedance analysis for the fundamental mode.} Bode plots and the corresponding Nyquist map for: (a) presence of a single type of generation sites. (b) presence of a single type of trapping sites. The calculated impedance in the absence of nonequilibrium carriers, $1/Z_{(1)}=G_0+i\omega C$, has been included in the Nyquist maps on the right panels as a continuous curve.} The \textit{apparent} circuit has been included on top.
\end{figure}

We will continue to acknowledge the symbolic appeal of these \textit{apparent} circuits but proceed to analyze the key features of the impedance per mode, focusing on Eqs.~\ref{adm1},~\ref{adm2}, and~\ref{adm3}, where substantial insights can be found.  
The frequency response will be depicted using Bode plots for a resistive like, $\text{Re}(1/Z_{(m)})$, and a reactive like component, $\text{Im}(1/Z_{(m)})\omega^{-1}$, complemented with the corresponding Nyquist's maps. We should note that it is common to find the expression, $\text{Im}(1/Z_{(m)})\omega^{-1}$, identified as an efective capacitance. For $m=1$, according to Eq.~\ref{adm1},
\begin{equation}
    \label{resist}
\text{Re}\left(\frac{1}{Z_{(1)}}\right)=\left( G_0+\frac{S}{2} \right)+\frac{S}{4}\frac{1}{1+(2\omega \tau)^2},
\end{equation}
and 
\begin{equation}
    \label{react}
\text{Im}\left(\frac{1}{Z_{(1)}}\right)\omega^{-1}=-\frac{S\tau}{4}\frac{1}{1+(2\omega \tau)^2}+C.
\end{equation}
This Lorenzian functional dependence in Eqs.~\ref{resist} and~\ref{react} characterizes the system ability to accept or release nonequilibrium carriers and is not related to a dielectric modulation.

The resulting impedance analysis for $m=1$
is depicted in Figures~\ref{Figure4} (a) and (b) for generation and trapping sites, respectively. The step-like behavior in the logarithmic frequency scale appears at $\omega \tau=1/2$ (half step height) that matches the condition for maximum hysteresis area reported in Ref.~\citenum{LopezRichard2022} for Type II  memristive responses. The height of the step, in turn, depends on the carrier mobility, the even component of the generation/trapping efficiency, $\sigma_{e}$, the relaxation time, and voltage amplitude, contained within $S$.
Given the dependence of $S$ on $\tau$,
the steps in the left panels scale as 
 $\text{Re}(1/Z_{(1)})\propto \tau$, while in the middle as $\text{Im}(1/Z_{(1)})\omega^{-1}\propto \tau^2$. More importantly, the reactive components in Eq.~\ref{react} can be positive or negative according to the sign of $\sigma_e$ and may even change sign if $C\neq 0$ resulting in the perceived  labels of negative capacitance or inductive contributions of nonmagnetic origin. This can also be identified in the corresponding Nyquist's maps, where arrows point to the frequency growth direction. These are very common figures found in metamaterials with memory traces. They have been labeled in the literature as traces of negative or positive capacitance~\cite{Jonscher1986,Ershov1998,MoraSero2006,Li2013,Ebadi2019} because of the resemblance to the semicircular $RC$ maps, or instead as inductive-like responses~\cite{Joshi2020,Priyadarshani2021,Gonzales2022}, for analogous reasons.
\begin{figure}[h]
	\includegraphics[width=8.5 cm]{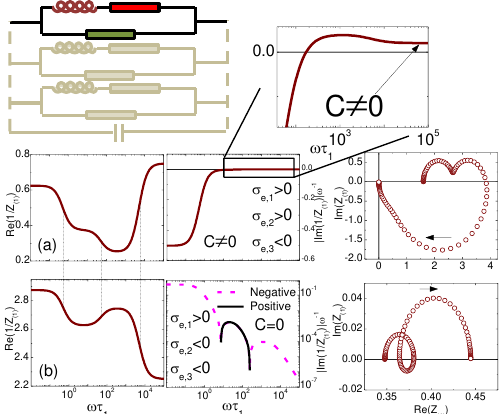}
	\caption{\label{Figure5} \textbf{Theoretical multi-mechanism approximation for the fundamental order mode.} Impedance analysis for the fundamental mode $m=1$ of three coexisting generation or trapping mechanisms with contrasting relaxation times. Bode plots and the corresponding Nyquist map for: (a) Combination of generation function components and a geometric capacitive coupling matching the experimental trend in Figures~\ref{Figure3} (c) and (d). (b) Removing the capacitance and changing the sign arrangement. The \textit{apparent} circuit has been included on top.}
\end{figure}

If the geometric capacitive coupling, whether intentional or not, is different from zero, the effect of the reactive part always vanishes in the limit of very high frequencies down or up to the geometric capacitive leftover, $C$, similar to our experiment in Figure~\ref{Figure3} (b) and a variety of other evidences as those reported for instance in Refs.~\citenum{Ershov1998,MoraSero2006,Joshi2020}. The corresponding Nyquist's plots exhibit also the interference of this capacitive effect, that has been singled out by plotting the impedance in the absence of nonequilibrium carriers, $1/Z_{(1)}=G_0+i\omega C$, as a continuous curve. However, a single step-like behavior cannot reproduce the non-monotonic experimental trend highlighted in Figure~\ref{Figure3} (a) and in the inset of panel (b).

While a single generation or trapping mechanism offers valuable insights, multiple concurrent processes, such as concomitant ionic and electronic transport, frequently occur. In the presence of additional independent mechanisms, when concurrent transport channels can be approximated as disconnected (e.g., in cases where electron-hole pair recombination is negligible during electron and hole transport, or where REDOX reactions are negligible during ionic drift), the contribution to the admittance can be considered additive. In such cases, the analysis can be easily extended, as outlined in the Methods Section.
This has been exemplified in Figure~\ref{Figure5} for $m=1$ considering three concomitant mechanisms with contrasting relaxation time scales, $\tau_{j+1}/\tau_{j}=10^{-2}$ combining generation and trapping processes. We have introduced logarithmic vertical axes in the middle panels of Figure~\ref{Figure5} since unlike the sequential steps in the left panels that scale as $\text{Re}(1/Z_{(1)})\propto \tau_{j}$, the steps in the middle panels scale as $\text{Im}(1/Z_{(1)})\omega^{-1}\propto \tau_{j}^2$ as discussed previously. Note, that we report the absolute value, $|\text{Im}(1/Z_{(m)})\omega^{-1}|$, in panel (b), where the function sign has been identified with contrasting lines. The non-monotonic trends of both resistive and reactive components and the sign of the latter emerge from the same nonequilibrium dynamics following the character of $\sigma_{e,j}$ according to the sequence of the relaxation times, $\tau_j$. Our interpretation differs from one that suggests the presence of capacitive and inductive effects.
The patterns in panel (a) can be recognized, in the experimental Figures~\ref{Figure3} (c) and (d) and also in the experiments reported in Refs~\citenum{Jonscher1986,Almora2015,Pivac2017,Guerrero2016,Gonzales2022}. Also the shape of the corresponding Nyquist's maps changes drastically according to the time-scale sequencing of the contributing mechanisms.

\begin{figure}[h]
	\includegraphics[width=8.5 cm]{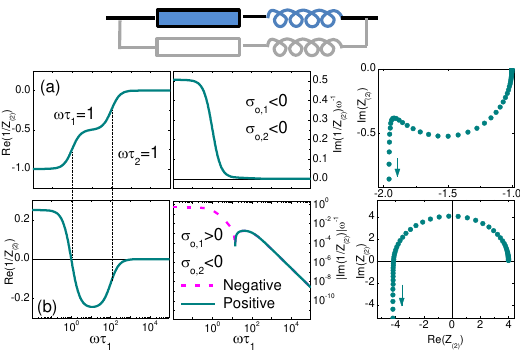}
	\caption{\label{Figure6} \textbf{Theoretical frequency domain multi-mode analysis for the second order mode.} Bode plots and the corresponding Nyquist's maps for $m=2$: (a) two coexisting mechanisms matching the experimental trend in Figures~\ref{Figure3} (e) and (f), (b) combining two generation function components with opposing signs. The \textit{apparent} circuit has been included on top.}
\end{figure}

Note that the contributions from Eq.~\ref{curr2}  proportional to the odd component of the generation rate, $\sigma_{o}$, (embedded in $A$), and all the information related to eventual inversion asymmetries and polarity would have been overlooked by just analyzing the fundamental mode, $m=1$. The effects of adding independent channels has been illustrated in Figures~\ref{Figure6} (a) and (b) for $m=2$. Again, a step-like behavior appears as the frequency is spanned across the marks, $\omega \tau_j=1$, for $j=1,2$ for charging mechanisms with the same $\sigma_o$ character, corresponding to the example in panel (a) that matches the experimental trend in Figures~\ref{Figure3} (e) and (f). Yet, changes in sign in panel (b) occur for alternating $\sigma_{o,j}$ sign during the sequential tuning of charging mechanisms with frequency. The resistive and reactive part of the impedance can be either positive or negative according to the nature of $\sigma_{o,j}$. For all the above cases the admittance for coexisting channels still collapses as $\omega \rightarrow \infty$. In this approximation, according to Eqs.~\ref{adm1} and \ref{adm3}, the information contained in the impedance of the mode $m=3$ could appear redundant to the one for $m=1$. However, one must point that while the fundamental mode, $m=1$, contains the interference of both $G_0$ and $C\neq 0$, the third order mode is fully independent on them, holding just the microscopic information enclosed within $S$. Yet, the susceptibility to errors inherent in the weak third-order mode turns this task challenging.

Note that throughout the text, we have used the term \textit{apparent circuits} rather than the more common \textit{equivalent circuits}. This distinction is important because, as shown in Figures~\ref{Figure2new} (a-c), it is not possible to fully capture the electrical response of devices with inherent memory using equivalent circuit projections based solely on the fundamental mode. An example of a circuit that can simultaneously account for both the memory response and impedance maps, as outlined by our model, is provided in the Supplementary Material.

\section[here]{Conclusions}

In summary, avoiding a multimode perspective when studying nonlinear transport responses may hinder a complete understanding of the causes and nature of the emergence of memory traces. It is thus recommended to try to match multimode impedance analyses with theoretical models starting with the simplest possible microscopic approach and gradually add complexity when necessary. As demonstrated here, conventional approaches can already provide relevant clues for seemingly complex and unrelated behaviors. 

This also highlights that, if one opts on building \textit{apparent} circuit representations starting from a microscopic perspective it is crucial to recognize the limitations in the interpretation associated to the \textit{apparent} circuit components. Following this perspective we are able to provide the actual \textit{apparent} circuit representation for a generic memristive response in the Supplementary Material. The emergence of negative capacitances or unexpected inductances (in the absence of any magnetic ingredients or eventual displacement currents) is a consequence of the apparent circuit selection.
Clearly, real capacitive effects, that lead to displacement currents, or inductive features, magnetic in nature, cannot be ruled out when characterizing transport properties. However, as we tried to prove here, the mere availability of nonequilibrium charges and the retardation of their path towards equilibrium are already sufficient ingredients to provide a very rich impedance response that matches a vast range of experimental evidences. 

\section[here]{Methods}

 The transport characteristics that emerges in nonlinear dynamic systems under alternating input, $V=V_{0} \cos{\omega t}$, at steady state, can only be reconstructed by a specific combination of higher order harmonics, as 
\begin{equation}
    \label{fourierm}
    I=V_0 \sum_{m} \left[G_{(m)} \cos\left( m \omega t\right) -  B_{(m)}\sin\left( m \omega t\right)\right].
\end{equation}
This multimode response allows defining an admittance, or inversed impedance, per mode, independent on time, 
\begin{equation}
    \label{admit}
    \frac{1}{Z_{(m)}(\omega)}= G_{(m)}(\omega)+i B_{(m)}(\omega).
\end{equation}
where $G_{(m)}$ can be interpreted as being the $m$-mode conductance and $B_{(m)}$, the $m$-mode susceptance. Such definitions emerge in compliance with the traditional way the spectral information is tackled in the impedance analysis for the fundamental (and only) mode, $m=1$, in linear responses.

When independent generation and/or trapping mechanisms contribute simultaneously, Eq.~\ref{taxa} transforms to
\begin{equation}
    \label{taxam}
    \frac{d \delta n_{j}}{dt}=-\frac{\delta n_{j}}{\tau_{j}} + g_{j}\left( V \right),
\end{equation}
with $g_{j}\left( V \right) \simeq \sigma_{o,j} V+\sigma_{e,j}V^2$ for small relative signals. Here, the index $j$ labels each independent transport mechanism, transforming the total current in Eq.~\ref{curr} as
\begin{equation}
    \label{currm}
    I=\left( G_{0} + \sum_{j} \gamma_{j} \delta n_{j} \right) V.
\end{equation}
The corresponding admittance per mode are thus given in this case by
\begin{equation}
    \centering
    \label{adm1m}
    \begin{aligned}
&\frac{1}{Z_{(1)}}=G_0+\sum_j \frac{S_{j}}{2}  \\
&+\sum_j\frac{S_{j}}{4}\left[\frac{1}{1+(2\omega \tau_j)^2}-i \frac{2\omega \tau_j}{1+(2\omega \tau_j)^2}   \right] \\
&+iC\omega,
\end{aligned}
\end{equation}
\begin{equation}
    \label{adm2m}
\frac{1}{Z_{(2)}}=\sum_jA_{j}\left[\frac{1}{1+(\omega \tau_j)^2}-i \frac{\omega \tau_j}{1+(\omega \tau_j)^2}   \right],
\end{equation}
and
\begin{equation}
    \label{adm3m}
\frac{1}{Z_{(3)}}=\sum_j\frac{S_{j}}{4}\left[\frac{1}{1+(2\omega \tau_j)^2}-i \frac{2\omega \tau_j}{1+(2\omega \tau_j)^2}   \right].
\end{equation}
Here $S_{j}=\gamma_{j}\sigma_{e,j}\tau_{j} V_0^2$ and $A_{j}=\gamma_{j}\sigma_{o,j}\tau_{j} V_0/2$.

The dynamic current-voltage measurements were conducted on the device reported in Ref.~\citenum{Miller2021} operating in a memristor configuration. The IS measurements were carried out using a Lock-In amplifier (EG\&G Instruments, Model: 7265) in current signal mode. The real and imaginary components of the current were measured on the actual device operating under floating gate configuration. This was achieved by applying a sinusoidal signal from the function generator and employing the same signal as a reference for the Lock-In amplifier.

\section[here]{Supplementary Material}

The supplementary material provides a detailed description of an apparent circuit model designed to replicate the behavior of a memristor under various conditions, with a particular emphasis on the small amplitude approximation regime. This model includes key components and parameters necessary for accurately reproducing memristive loops across different scenarios.

\section[here]{Acknowledgments}

This study was financed in part by the Conselho Nacional de Desenvolvimento Científico e Tecnológico - Brazil (CNPq) Projs. 311536/2022-0 and 131951/2021-1, FAPESP Proj. 2023/05436-3, and the Fulbright Program of the United States Department of State’s Bureau of Educational and Cultural Affairs. 
We are furthermore very grateful to  B. Leikert, J. Gabel M. Sing and R. Claessen from Physikalisches Institut, Experimentelle Physik 4, Universität Würzburg, 97074 Würzburg, Germany for the sample growth of the devices used in this study.

\section[here]{Data Availability Statement}

The data that support the findings of this study are available from the corresponding author upon reasonable request.


\bibliography{bibliography}
\end{document}